\documentclass[preprint,12pt]{elsarticle}

\usepackage{amssymb}
\journal{Journal of Computational Physics}

\begin{document}

\begin{frontmatter}



\title{Numerical studies of light-matter interaction driven by plasmonic fields: the velocity gauge}


\author[icfo]{A. Chac\'on}
\author[icfo,icrea]{M. Lewenstein}
\author[mpq]{M. F. Ciappina}

\address[icfo]{ICFO-Institut de Ci\`ences Fot\`oniques, Av. Carl Friedrich Gauss 3, 08860 Castelldefels (Barcelona), Spain}
\address[icrea]{ICREA-Instituci\'o Catalana de Recerca i Estudis Avan\c{c}ats, Lluis Companys 23, 08010 Barcelona, Spain}
\address[mpq]{Max-Planck-Institut f\"ur Quantenoptik, Hans-Kopfermann-Str. 1, 85748 Garching, Germany}

\begin{abstract}
Theoretical approaches to strong field phenomena driven by plasmonic fields are based on the 
length gauge formulation of the laser-matter coupling.  
From the theoretical viewpoint it is known there exists no preferable gauge and 
consequently the predictions and outcomes should be independent of this choice. The use of the 
length gauge is mainly due to the fact that the quantity obtained from finite elements simulations of plasmonic fields is the {\it plasmonic enhanced laser electric field} rather than the laser vector potential. In this paper we develop, from first principles, 
the velocity gauge formulation of the problem and we 
apply it to the high-order harmonic generation (HHG) in atoms. A comparison to the results obtained with the length gauge is made. It is analytically and numerically
demonstrated that both gauges give equivalent descriptions of the emitted HHG spectra resulting from the interaction of a spatially inhomogeneous field and the single active electron (SAE) model of the helium atom. We discuss, however,  advantages and disadvantages of using different gauges 
in terms of numerical efficiency. 
  
\end{abstract}


\end{frontmatter}


\section{Introduction}
\label{intro}

Nowadays there exists a high demand for coherent light sources extending from the ultraviolet (UV) 
to the extreme ultraviolet (XUV) spectral ranges. These sources provide important tools for basic research, material science and biology among other branches~\cite{ferencreview}. An important obstacle preventing these sources from reaching high efficiency and large duty cycles is  their demanding infrastructure. The recent demonstration of XUV generation driven by surface plasmon resonances, conceived as {\it light enhancers}, could provide a plausible 
solution to this problem~\cite{kim}. The high-order-harmonic generation (HHG) in atoms 
using plasmonics fields, generated starting from tailored metal nanostructures,
requires no extra amplification of the incoming pulse. 
By exploiting the so-called surface plasmon polaritons (SPP),
the local electric fields can be enhanced by several orders of magnitude~\cite{kim,park,pfullmann}, 
thus exceeding the threshold laser intensity for HHG generation in noble gases. 
One additional advantage is that the pulse repetition rate remains unaltered without any extra pumping or
cavity attachment. Furthermore, the high-harmonics radiation, 
generated from each nanostructure typically in the UV to XUV range, acts as a source with point-like properties, 
enabling collimation or focusing of this coherent radiation by
means of constructive interference. This opens a wide range of possibilities to spatially arrange 
nanostructures to enhance or shape the spectral and spatial 
properties of the outgoing coherent radiation in numerous ways.

One  can shortly describe  the high-order-harmonic generation based 
on plasmonics fields as follows (a more exhaustive 
description can be found in the seminal paper of Kim et al.~\cite{kim}): a femtosecond low-intensity laser 
pulse is coupled to the plasmon mode of a metal nanostructure inducing
a collective oscillation of the free electrons within the metal. These free charges redistribute the electric 
field of the laser around each of the nanostructures, thereby forming a spot of highly enhanced electric field, 
also known as {\it hot spot}. 
The plasmon amplified field exceeds the threshold of HHG, thus by injection of a gas jet, 
typically a noble gas, onto the spot of the enhanced field, high order harmonics from the gas atoms are generated.
In the original experiment of Kim et al.~\cite{kim}, the output laser beam emitted from a low-power femtosecond oscillator was directly focused onto a
$10\times10$ $\mu$m size array of bow-tie nanoantennas with a pulse intensity
of the order of 10$^{11}$ W/cm$^{2}$, which is about two orders of magnitude
smaller than the threshold intensity to generate HHG in noble gas atoms.
The experimental result of Ref.~\cite{kim} showed that the field intensity
enhancement factor exceeded 20 dB, i.e. the enhanced laser intensity is two orders
of magnitude larger than the input one, which is enough to
produce from the 7th to the 21st harmonics of the fundamental frequency by injecting xenon gas. 
For the case of  the laser wavelength corresponding to a Ti:Sa laser, i.e.~about 800 nm, the wavelength of the emitted
coherent radiation is between 38 nm and 114 nm. Additionally, each bow-tie nanostructure acts as a
point-like source, thus a three-dimensional (3D) arrangement of bow-ties should enable us to perform 
control of the properties of generated harmonics, e.g.~their polarization, in various ways by exploiting 
interference effects. Due to the strong confinement of the plasmonic hot spots, which are of nanometer size, the 
laser electric field is clearly no longer spatially homogeneous in this tiny region. Since typically 
electron excursions are of the same order as the size of this region, important changes in the laser-matter processes occur, see. e.g.~\cite{ciappi2012,ciappioptexp, joseprl}. 

So far, all of the the numerical approaches to study laser-matter processes in atoms and molecules driven
by plasmonic fields, in particular HHG and ATI,
are based on the length gauge of the laser-coupling formulation~\cite{husakou2011,yavuz2012,tahir2012,ciappi2012a,tahir2012a,tahir2013,fetic,tahir2013a,tahir2013b,yavuz2013,ciappi2013,luo2013,ciappi2013lpl,ciappi2014,feng2013,ciappicpc,ciappijpcs}.  
The use of the length gauge is mainly due to the fact that the quantity obtained from finite elements simulations of plasmonic fields is the {\it plasmonic enhanced laser electric field} rather than the laser vector potential. Only a couple of papers employed 
an extension of the Strong Field Approximation (SFA), where an approximate version of the velocity 
gauge was used~\cite{ciappi2012,tahir2013b}.  Different descriptions of light 
matter interaction (c.f. Ref.~\cite{morten}), which include the full spatial dependence 
of the electromagnetic field, are closely related to 
the problem presented in our contribution.
There are, however, distinct differences 
amongst the general formulation of the non-dipole problem with
 the one we will tackle in the present article. For instance, the next 
 order of the non-dipole description includes both the electric quadrople 
 and the magnetic dipole terms, which are not present in our plasmonic 
 fields, because the typical laser intensities are far below the ones
  needed to consider relevant these effects.

In this article, we concentrate our effort on the formulation and 
numerical implementation of the velocity 
gauge description of light-matter interaction driven by plasmonic 
fields. From a pure theoretical viewpoint, it is known 
the velocity gauge is more appropriate and consequently 
our contribution will fill the missing gap, completing the 
whole picture in the modeling of laser-matter processes driven by plasmonic fields.

The paper is organized as follows. In Sec. II, we shall present 
the velocity gauge formulation of the problem and 
we relate it to the length gauge, clearly showing
the compatibility between them. The numerical implementation
 is presented in Sec. III, joint with a set of
examples and a discussion about how the 
two different algorithms, i.e. the spectra split operator and the Crank-Nicolson,
behave as a function of the relevant parameters. 
Furthermore an analysis of the
computational efficiency and scaling of both formulations is presented 
here. The paper ends with a short summary and an outlook.


\section{Theory and gauge transformation} 
Quantum mechanics governs the evolution of the systems, 
atoms and molecules in our case, when they interact with an extental electromagnetic field.  
In particular, the Time Dependent Schr{\"o}dinger Equation (TDSE)~\cite{Sakurai1994} allows us to 
obtain the complete time-space evolution of the particles. From a mathematical viewpoint,
  there are two different, but equivalent, expressions for the Hamiltonian 
which describes the interactions of the whole system.  As a consequence the laser-matter 
problem can be formulated both in the so called velocity gauge (VG) or in 
the length gauge (LG), indistinctly. Formally, both gauges present 
equivalent descriptions of the quantum problem~\cite{Sakurai1994}, 
and therefore the results should not change if either the 
VG or LG is utilized to compute the observables of interest. 
Here, we detail how the gauge transformation is commonly 
implemented in the laser-matter interaction and in particularly 
when a spatial inhomogeneous field interacts with an atomic or molecular target. 
In general, we are interested in to describe the electron dynamics of an atomic 
or molecular system when it interacts with an electromagnetic field. For this case the TDSE reads 
(atomic units are used throughout the paper otherwise stated): 
\begin{eqnarray}
\hspace{0cm}H\Psi({\bf r},t) &=& i\frac{\partial}{\partial t}\Psi({\bf r},t), \label{TDSE}
\end{eqnarray}
where, $H$, is the Hamiltonian of the quantum system and $\Psi({\bf r},t)$ is the electron wavefunction (EWF).

Let us define the Hamiltonian, $H_{\rm V}$, in the minimum coupling or VG
for the electromagnetic field-matter interaction as: 
\begin{eqnarray}
\hspace{0cm}H_{\rm V} &=& \frac{1}{2}\left[{{\bf p}+{\bf A}({\bf r},t)}\right]^2 + V_0({\bf r}), \label{Vgauge}
\end{eqnarray}
where, ${\bf p}=-i\nabla$, denotes the {\it canonical momentum} operator, 
${\bf A}({\bf r},t)$, is the vector potential of the electromagnetic field, which in this case 
corresponds to a spatial inhomogeneous or plasmonic field. In Eq.~(\ref{Vgauge}),
$V_0({\bf r},t)$ is the electrostatic Coulomb interaction between the charged particles. 
The vector potential for the spatial inhomogeneous field typically can be represented in the following form:
\begin{eqnarray} 
 {\bf A}({\bf r},t)&=&[1+\epsilon g({\bf r})]{\bf A}_{\rm h}(t),\nonumber \\  
 {\bf A}_{\rm h}(t)&=& A_0 f(t)\sin(\omega_0 t + \varphi_{\rm CEP}){{\bf e}_z}. \label{VPot} 
\end{eqnarray}
Here, ${\bf A}_{\rm h}(t)$, denotes the homogeneous or conventional vector potential, $A_0$ 
is the amplitude of the vector potential, $\omega_0$, is the central frequency,
$\varphi_{\rm CEP}$ is the carrier-envelope-phase (CEP) and $f(t)$ is a function which defines the 
time envelope of the field. $\epsilon$ is a small parameter that governs 
the strength of the spatial inhomogeneity (see e.g.~\cite{ciappi2012} for more details) 
and $g(\bf{r})$ describe the spatial dependence of the plasmonic field. 
Note that in the limit when $\epsilon=0$, the vector potential 
field does not depend on the spatial coordinate anymore and we recover
 the conventional laser-matter formulation. The units of $\epsilon$ depend 
 on the function $g(\bf{r})$. For instance, if $g({\bf r})=z$ (a linear function), $\epsilon$ 
 has units of inverse length (see e.g.~\cite{ciappi2012}).

Often, it is desirable to solve the TDSE in the length gauge or 
the maximal coupling gauge. This is so because the numerical or analytical 
calculation can be expressed in an easy way and the computation 
of certain observables is more efficient~\cite{Cormier1996}. 
Therefore, the main question is how we can perform the transformation 
of the Hamiltonian in the VG, Eq.~(\ref{Vgauge}), to the LG. 
The gauge transformation should be boiling down in an unitary translation of the 
whole wavefunction~\cite{GT0}. We define this unitary transformation according to:
\begin{eqnarray}
\hspace{0cm}\Psi_{\rm V} &=&  Q^\dagger\Psi_{\rm L}, \label{WF_VG} 
\end{eqnarray}
where, $\Psi_{\rm V}=\Psi_{\rm V}({\bf r},t)$ and $\Psi_{\rm L}=\Psi_{\rm L}({\bf r},t)$ 
are the wavefunctions in the VG and LG, respectively. $Q$ is the unitary hermitian operator defined 
according to the following rule $Q=\exp{[i\chi({\bf r},t)]}$~\cite{GT0,Cohen1977,Sakurai1994}, with
$\chi({\bf r},t)=\int_{\rm C}^{{\bf r}} {\bf A}({\bf r}',t)\cdot d{{\bf r}'}$. The latter expression is a contour integral
which is independent of the path, because we can safely assume that the effect of the magnetic field is negligible, i.e.~that the curl of the vector
 potential for the inhomogeneous field is zero, $\nabla \times {\bf A}={\bf 0}$.  
Furthermore, by using Eqs.~(\ref{TDSE}) and (\ref{WF_VG}), we find 
the transformation for the Hamiltonian, $H_{\rm V}$, from the VG to the LG:
\begin{eqnarray}
QH_{\rm V} Q^\dagger\Psi_{\rm L} &=&  \frac{\partial{\chi({\bf r},t)}}{\partial t} \Psi_{\rm L} + i\frac{\partial\Psi_{\rm L}}{\partial t}. \label{Der0G} 
\end{eqnarray}

Then, knowing that ${\bf E}({\bf r},t) = -\frac{\partial }{\partial t}{\bf A}({\bf r},t)$,
 i.e.~the relationship between ${\bf E}({\bf r},t)$ and {\bf A}({\bf r},t), the last expression becomes:
\begin{eqnarray}
\hspace{0cm}\left[QH_{\rm V} Q^\dagger+ \int_{\rm C}^{{\bf r}}{\bf E}({\bf r}',t)\cdot d{{\bf r}'}\right] \Psi_{\rm L} &=&   i\frac{\partial\Psi_{\rm L}}{\partial t}. \label{Der0G1} 
\end{eqnarray}
As the TDSE is gauge invariant, we infer that the Hamiltonian in the VG, $H_{\rm V}$, 
is transformed to the LG, $H_{\rm L}$, via:
\begin{eqnarray}
\hspace{0cm}H_{\rm L} &=&QH_{\rm V} Q^\dagger+ \int_{\rm C}^{{\bf r}}{\bf E}({\bf r}',t)\cdot d{{\bf r}'}. \label{Der0G2} 
\end{eqnarray}
It can be demonstrated that the first term on the right hand side of  Eq.~(\ref{Der0G2}), 
yields $QH_{\rm V} Q^\dagger = \frac{1}{2}{\bf p}^2 + V_0({\bf r})$. 
Then, the Hamiltonian in the LG takes the form: 
\begin{eqnarray}
\hspace{0cm}H_{\rm L} &=&\frac{{\bf p}^2}{2}+V_0({\bf r})+ V_{\rm int}({\bf r},t), \label{HamLG0} 
\end{eqnarray}
here, ${\bf p}$ is the {\it kinetic momentum} operator, and $V_{\rm int}({\bf r},t)= \int_{\rm C}^{{\bf r}}{\bf E}({\bf r}',t)\cdot d{{\bf r}'}$, 
is a contour integral.  In terms of Classical Mechanics, we can interpret this last term
as the work done in the electric field ${\bf E}({\bf r}',t)$ to move the electron 
from an arbitrary place to the position $\bf{r}$. In the particular case when the vector 
potential has the functional form given by Eq.~(\ref{VPot}) and the function $g({\bf r})$, is 
set to $g({\bf r})=z$, the Hamiltonian operator in the LG becomes:
\begin{eqnarray}
\hspace{0cm}H_{\rm L} &=&\frac{{\bf p}^2}{2}+V_0({\bf r})+ z(1+\frac{\epsilon}{2}z)E_{\rm h}(t), \label{HamLG1} 
\end{eqnarray}
where, $E_{\rm h}(t) = -\frac{\partial }{\partial t}A_{\rm h}(t)$ denotes the spatial homogeneous part of the laser electric field. Commonly this field, $E_{\rm h}(t)$, is called the {\em conventional} or spatial homogeneous field. 

In the next section, we shall compare the numerical accuracy of the VG and LG predictions 
for the high-order harmonic generation (HHG) driven by plasmonic fields. Our numerical models 
are based on Eqs.~(\ref{Vgauge}), for the VG, and (\ref{HamLG1}), for the LG, respectively.

\section{Numerical algorithms} 

The methods utilized to numerically integrate the TDSE are classified by considering 
how the time evolution of the EWF is computed. When the EWF at a later time is obtained 
from the one at the current time, we have the so-called explicit methods. On the other hand, 
implicit schemes find the EWF by solving an equation involving both the actual EWF and 
one at later time. We choose the Spectral-Split Operator (SO) method joint with
the Crank-Nicolson (CN) scheme, which are both explicit methods, to numerically 
integrate the TDSE of our interest. The SO uses a spectral technique to evaluate 
the derivative operator in the Fourier domain~\cite{Feit1982,NumericalRecipe}, and, 
on the other hand, the CN is based on the finite element difference discretization 
technique~\cite{NumericalRecipe} to implement the second derivative present in the  
Hamiltonian, which defines the kinetic operator term.  

\noindent In order to test the accuracy of both the VG and the LG in the HHG driven by 
plasmonic fields, we have implemented the TDSE via the SO and CN techniques within a one 
spatial dimension model (1D).  

\noindent A general solution of the TDSE is done by employing a unitary $U(t_0+\Delta t,t_0)$ 
evolution operator, where $t_0$ is the initial time, i.e.~the initial EWF $\Psi_0(t_0)$ is 
known and we evolve the system to an unknown state $\Psi(t_0+\Delta t)$ 
at a given time $t_0+\Delta t$~\cite{Sakurai1994}: 
 \begin{eqnarray}
\Psi(t_0+\Delta t) &=& U(t_0+\Delta t,t_0)\Psi_0(t_0). \label{TimeEvol}
\end{eqnarray}
For simplicity, in Eq.~(\ref{TimeEvol}), we have dropped out the spatial (${\bf r}$) dependence on the EWF. 
In the laser-matter community, the $U(t_0+\Delta t,t_0)$ is commonly known as a 
propagator and it has the following explicit form, $U(t_0+\Delta t,t_0)=\exp\left[-i\int_{t_0}^{t_0+\Delta t}H(t')dt'\right]$. 

\noindent In reference~\cite{Feit1982}, Feit et al.~have introduced the SO method to numerically solve
the TDSE in two spatial dimensions (2D) by using Eq.~(\ref{TimeEvol}). This method consists 
in to split the time evolution operator $U(t_0+\Delta t,t_0) \approx e^{-iH(t_0+\frac{\Delta t}{2})\Delta t}$ 
in three parts~\cite{Feit1982}: 
 \begin{eqnarray}
\hspace{0cm}\Psi(t_0+\Delta t) &=& e^{-i\frac{1}{2}{\bf p}^2{\Delta t/2}}e^{-iV_{\rm eff}(t_0+\frac{\Delta t}{2}){\Delta t}}e^{-i\frac{1}{2}{\bf p}^2{\Delta t/2}}\Psi_0(t_0). \label{TimeEvol1}
\end{eqnarray} 
Here, the Hamiltonian, $H(t_0+\frac{\Delta t}{2})$, is divided in 
$H(t_0+\frac{\Delta t}{2})= \frac{1}{2}{\bf p}^2 + V_{\rm eff}(t_0+\frac{\Delta t}{2})$, 
with $V_{\rm eff}(t) = V_0({\bf r})+\int_{\rm C}^{\bf r}{\bf E}({\bf r}',t)\cdot d{\bf r}'$ 
the effective potential in the LG. The main advantage of Eq.~(\ref{TimeEvol1}) is that we can evaluate the kinetic operator term, $e^{-i\frac{1}{2}{\bf p}^2 {\Delta t}/2} \Psi_0({\bf r},t_0)$, acting on the
 initial state, in the momentum space. This means that we need to compute a Forward Fourier 
 Transform (FFT)~\cite{FFTW} 
 of $\Psi_0({\bf r},t_0)$ and then multiply it by a phase factor which evaluates the action 
 of the kinetic operator, instead of a complicated derivate operator. Then, over this momentum 
 space EWF, an Inverse Fourier Transform (IFT) is applied in order to return to the coordinate 
 space~\cite{Feit1982}. This procedure is performed because the momentum 
 operator in the conjugate (momentum) space is just a number and not a derivative one. 
  
\noindent For the conventional or homogeneous fields case, the vector potential 
does not depend on the spatial coordinate, i.e.~${\bf A}({\bf r},t)={\bf A}(t)$, allowing 
us to  evaluate the kinetic operator in the VG as: 
 \begin{eqnarray}
\hspace{-.0cm}\Psi(t_0+\Delta t) = e^{-i\frac{1}{2}({\bf p}+{\bf A}(t_0+\Delta t/2))^2{\Delta t/2}}e^{-iV_{0}{\Delta t}}e^{-i\frac{1}{2}{({\bf p}+{\bf A}(t_0+\Delta t/2))}^2{\Delta t/2}}\Psi_0(t_0). \label{TimeEvol2}
\end{eqnarray} 
Clearly, this is not the case for the spatial nonhomogeneous fields. 
The dependence of the vector potential on the position, as stated in
 Eq.~(\ref{VPot}), does not allow us to apply Eq.~(\ref{TimeEvol2}). 
 This is so because in the momentum space the position operator
  becomes a derivative, which complicates substantially the SO method. 
Therefore, we conclude that the SO method can not be easily employed
 to numerically integrate the TDSE in the VG. However, by using a finite 
 element grid discretization, we will show that the CN method can be used
  in both gauges, VG and LG. The CN method is based on the solution 
  of Eq.~(\ref{TimeEvol}) by the Caley formula and the evaluation of the 
  kinetic operator in the position space using a finite element
   method~\cite{NumericalRecipe}. In 1D the numerical algorithm can be written as: 
\begin{eqnarray}
\left[{1+i\frac{\Delta t}{2}H(t_0+\Delta t/2)}\right]\Psi(t_0+\Delta t) = \left[{1 -i\frac{\Delta t}{2}H(t_0+\Delta t/2)}\right]\Psi_0(t_0). \label{TimeEvol3}
\end{eqnarray}  
The unknown EWF, $\Psi(t_0+\Delta t)$, is then computed by solving a tridiagonal system of equations.

\section{System description  and results}
In Attosecond Science, high-order harmonic generation (HHG) is one of the most important phenomena. 
For instance,  it is possible to synthesize attosecond pulses or to obtain structural information about the atomic of molecular systems~\cite{ferencreview} from the HHG spectra. Therefore, we chose here this observable driven by conventional (homogeneous) 
and non-homogeneous fields to compare the accuracy of our VG and LG implementations. 

For simplicity, we restrict ourselves to a one dimensional (1D) model, although it is known this approach is able to accurately reproduce the main features of the HHG spectra of real atoms~\cite{knightreview}. The potential well, $V_0(x)$, which defines our atomic system, is a soft-core or quasi Coulomb potential:
\begin{eqnarray}
V_0(x) &=& -\frac{Z}{\sqrt{x^2+a}},
\end{eqnarray} 
where $Z$ is the atomic charge and $a$ a parameter which allows us to tune the ionization potential of the atom of interest. In this paper, we set
$Z=1$ and $a=0.488$ a.u., such as the ionization potential is $I_p=0.9$~a.u. ($24.6$~eV), 
i.e the value for the single active electron (SAE) 
model of the He atom~\cite{HeliumSAE}.  Our ground state was computed 
via imaginary time propagation for a different set of spatial grid steps $\delta x$. To assure 
a ``good time" step, $\delta t$, convergence, we have used the criterion:
 $\delta t < {\delta x^2}/{2}$ (for more details see e.g.~\cite{NumericalRecipe}). 

In order to compute the HHG spectra, we firstly calculate the 
dipole acceleration expectation value, $a_d(t)$, as a function of time:
\begin{eqnarray}
a_d(t) &=& \langle\Psi(t)|\frac{\partial V_0(x)}{\partial x}+E(x,t)|\Psi(t)\rangle, \label{DipAccel}
\end{eqnarray} 
where the EWF $\Psi(x,t)$ is obtained via the SO and CN methods already described 
in the previous Section. The spectral intensity, $I_{\rm HHG}(\omega) = |\tilde{a}_d(\omega)|^2$, 
of the harmonic emission is then computed by Fourier transforming the dipole acceleration by using: 
\begin{eqnarray}
\tilde{a}_d(\omega) &=& \int_{-\infty}^{+\infty}{dt'a_d(t')e^{i\omega t'}}. \label{IntHHG}
\end{eqnarray}
The numerical computation of the HHG spectra will be performed by using a set of position steps $\delta x=\{0.05,~0.1,~0.15,...\}$~a.u. 
Consequently, and in order to estimate the numerical convergence of the HHG spectra as a function of $\delta x$, 
we use the spectral intensity difference between the smallest step, i.e. $\delta x_0=0.05$~a.u., 
and the rest of the set, 
$\Delta I_{\rm HHG,\delta x_j}(\omega)=|I_{\rm HHG,\delta x_0}(\omega)-I_{\rm HHG,\delta x_j}(\omega)|$, 
with $j=\{1,2,...\}$. Furthermore, for each of the $\delta x$, 
the computing time is also retrieved for both the VG and LG.

\subsection{HHG driven by conventional fields}\label{cap:System}
Firstly, we present computations of the harmonic spectra intensity, $I_{\rm HHG}(\omega)$, driven by a conventional homogeneous field. This case is the limit  $\epsilon\rightarrow 0$. Both numerical methods above described, i.e.~the SO and CN, have been used to compute the emitted harmonic spectra intensity both in the VG and LG. We shall show below that both gauges give the same results. 

\begin{figure}
\begin{center}
\resizebox{1\textwidth}{!}{\includegraphics{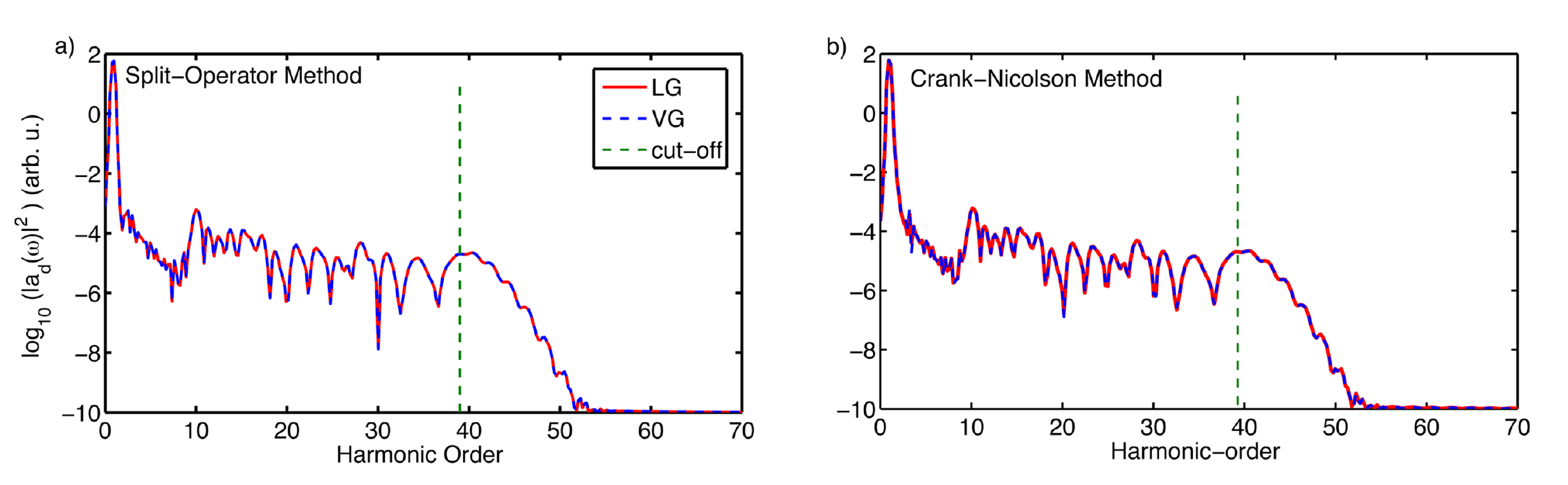}}
\caption{\label{HomoField_VG_LG_CN_SO} (color online) Computed high-order harmonic intensity spectra
 (in arbitrary units) driven by a spatial homogeneous (conventional) laser field under the LG (red solid line) and the VG (blue dashed line). Panel (a) the HHG spectra are obtained by using the SO method, panel (b) the same as (a) but using the CN method. The green vertical dashed line depicts the classical harmonic cut-off law, i.e., $n_c= (I_p+3.17\, U_p)/\omega_0$~\cite{maciej}. The laser pulse parameters for these simulations are:  intensity $I_0=2\times10^{14}$~W/cm$^2$, carrier frequency $\omega_0=0.057$ a.u. (corresponding to a wavelength of $\lambda=800$ nm), and CEP, $\varphi_{\rm CEP}=0$ rad. The pulse envelope is a ${\sin}^2$ function with four total cycles. We chose a grid step of $\delta x=0.05$ a.u. for both gauges.}
\end{center}
\end{figure}

The TDSE calculations are performed in a grid with a step $\delta x=0.05$ a.u., and a spatial grid length of $L_x=3500$ a.u. The real-time evolution is done with a time-step of $\delta t=0.00125$ a.u. Fig.~\ref{HomoField_VG_LG_CN_SO} shows the spectral intensity of the harmonic emission when a laser pulse interacts with our 1D helium model. In Fig.~1(a), the comparison of the HHG spectra between the LG and VG is depicted by using the SO method. The same comparison is shown in Fig.~1(b), but here the CN method is used for the numerical integration of the TDSE. 
Both methods show a perfect agreement when the LG and VG are used to compute the spectral harmonic intensity. This confirms that our numerical methods are able to describe the HHG process for any of the grid steps used in our simulations. 

\begin{figure}
\begin{center}
\resizebox{1.00\textwidth}{!}{\includegraphics{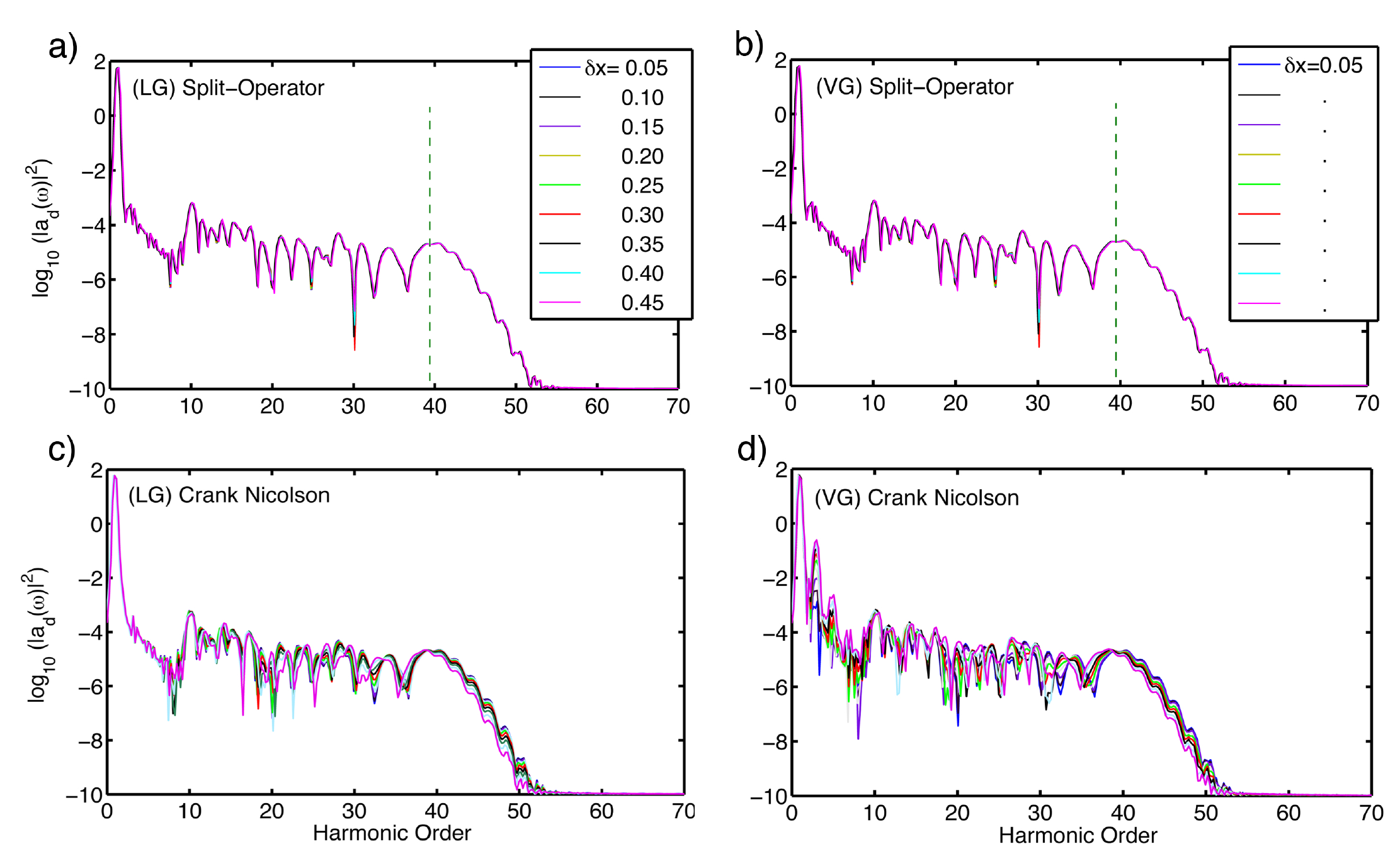}}
\caption{\label{HomoFieldVG_LG_CN} (color online) Computed high-order harmonic intensity spectra driven by a spatial homogeneous
  (conventional) field (in arbitrary units) as a function of the grid step, $\delta x$, under the LG and the VG by the SO method, panels (a)-(b) and the CN method panels (c)-(d). The green vertical dashed line depicts the classical harmonic cut-off law. The laser pulse parameters for these simulations are the same as in Fig.~1, i.e. intensity $I_0=2\times10^{14}$~W/cm$^2$, carrier frequency $\omega_0=0.057$ a.u. (corresponding to a wavelength of $\lambda=800$ nm), and CEP, $\varphi_{\rm CEP}=0$ rad. The pulse envelope is a ${\sin}^2$ function with four total cycles. }
\end{center}
\end{figure}

\noindent As a next test, we have integrated the TDSE for a set of grid steps,
 $\delta x=\{0.05,\,0.15,\,...,0.45\}$~a.u., and computed the emitted harmonics. 
 Figure~\ref{HomoFieldVG_LG_CN}, shows the results of the harmonic intensity, 
 $I_{\rm HHG}(\omega)$, as a function of the grid step computed by the  
SO Figs.~\ref{HomoFieldVG_LG_CN}(a)-(b) and the CN 
Figs.~\ref{HomoFieldVG_LG_CN}(c)-(d) methods, respectively.  
For both the VG and LG, the numerical spectra by using the SO 
method shows a perfect agreement for all the set of grid steps, $\delta x$ 
used in our simulations. In contrast, the situation is different when the CN 
method is employed to compute the harmonic spectra. 
For instance, the Figs.~\ref{HomoFieldVG_LG_CN}(c)-(d) show that the emitted harmonic spectrum 
depends on the grid step when the LG and VG are employed to computed the HHG. 
In addition, the computed HHG spectra slightly differ in the whole harmonic-order range whether the LG or the VG is 
used in the calculation and for the larger grid steps, i.e., $\delta x \geq 0.25$~a.u. Additional structures can be observed in the low-order harmonics for the case of VG (see Figs.~\ref{HomoFieldVG_LG_CN}(d)), although the general shape, including the harmonic cutoff, is in excellent agreement with the rest of the schemes. Considering the numerical error that the finite element method has
for the second derivative as a function of the grid spacing $\delta x$, it is reasonable 
to attribute poor convergence when the CN method is used with larger grid steps $\delta x$. Furthermore, in view of the fact that the 
VG has an extra spatial derivative of first order within the Hamiltonian, ${\bf p}\cdot{\bf A}$, 
we would expect that the numerical accuracy decreases when the spatial step, $\delta x$, 
increases. This is the reason behind the noticeable difference between the LG and the 
VG when larger grid steps are employed in the calculations of the HHG. Our numerical 
results show, however, that this difference between LG and VG 
disappears for the smallest spatial grid steps, i.e., $\delta x \leq 0.2$~a.u. 
These outcomes suggest that the best method to 
compute the HHG spectrum is the SO. On the other hand, in cases where the SO method is challenging, the
 CN method can be used if the grid step is small enough, e.g. $\delta x \lesssim 0.1$~a.u. We should
 note that the grid step will depend on the particular problem, i.e.~laser parameters, etc., although we can
  expect a general trend. For this reason, we suggest to perform a convergence analysis if the CN method is
  employed and to chose the adequate parameters
 for the required accuracy.

In the next, we shall perform the computation of the HHG spectra driven by a spatial inhomogeneous field. 
For the reasons explained in Section 3, we shall only use the CN method and compute the harmonic emission
 both in the LG and VG. 

\subsection{HHG driven by spatial inhomogeneous fields}
As was mentioned at the outset, when a laser field is focused on a metallic nanostructure, a hot spot of higher
 intensity, high enough to exceed the threshold for HHG in atoms, is created due to the coupling between the
  incoming field and the surface plasmon polaritons (SPPs)~\cite{kim}. The main property of the effective laser electric field is
   that it presents a spatial variation in the same scale as the one of the dynamics of the active electron.
Therefore, the interaction between this plamonic field and the atomic electron, which governs the HHG
 process, will change substantially. As the electric field is not
anymore spatially homogeneous, the electron will experience different electric field
 strengths along its trajectory. The question that emerges is which gauge can give us a numerical 
advantage when the TDSE is solved for the computation of the HHG spectra driven 
by spatial inhomogeneous fields. 
Before to address this question, we firstly demonstrate that both the LG and 
VG are equivalent in the description of the HHG driven by nonhomogeneous 
fields, as was demonstrated by the conventional case (see Section 4.1).

\begin{figure}
\begin{center}
\resizebox{0.80\textwidth}{!}{\includegraphics{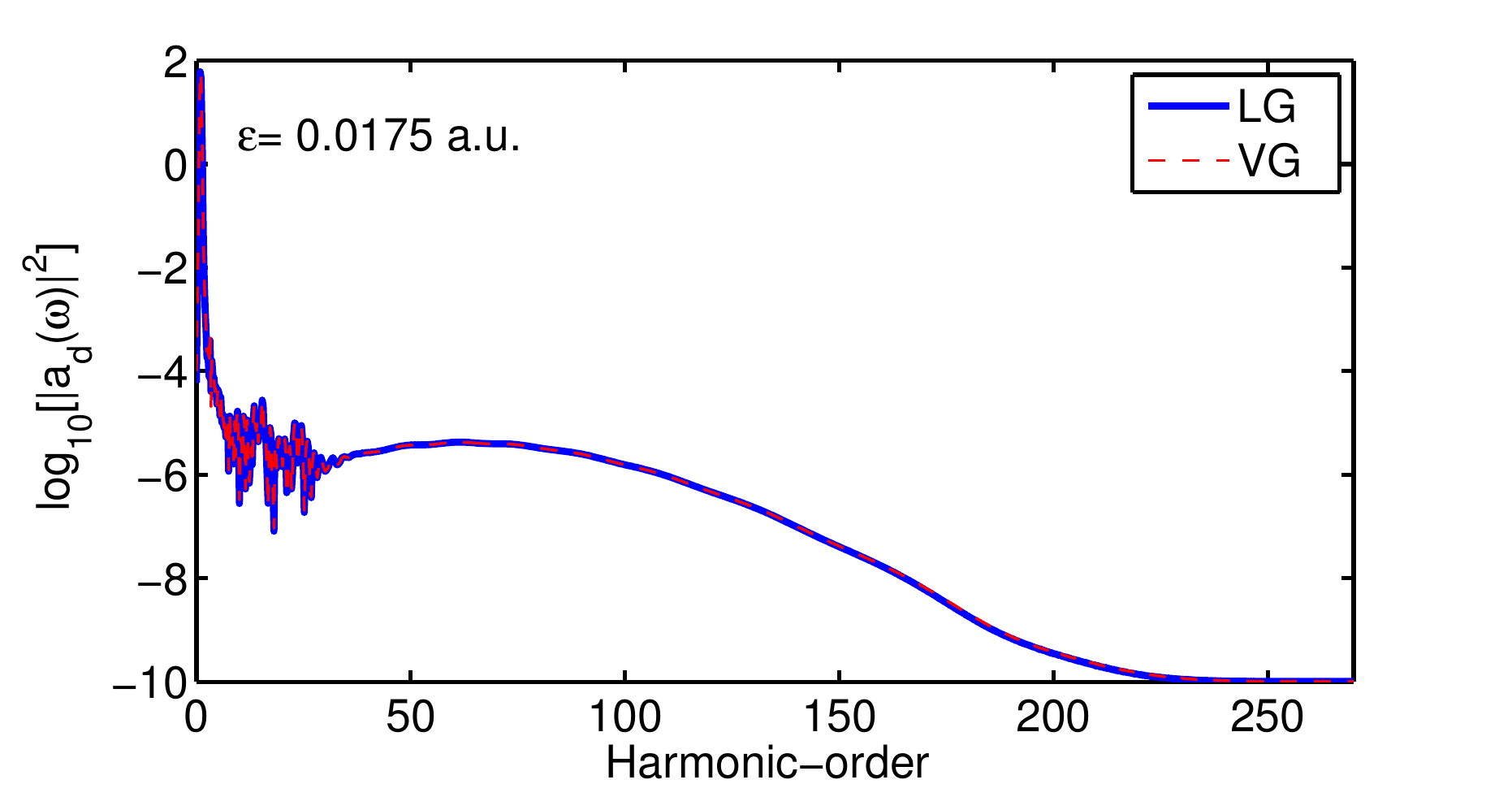}}
\caption{\label{inHomoFieldVG_LG_CN} (color online) HHG spectra driven by a spatial inhomogeneous field computed by using the LG (blue line) and VG (red dashed line).  The parameters for the laser pulse are the same that those used in Fig.~\ref{HomoFieldVG_LG_CN},  the inhomogeneous parameter is $\epsilon=0.0175$ a.u. (see the text for more details) and the grid step is $\delta x=0.05$~a.u.}
\end{center}
\end{figure}

\noindent We have numerically integrated the TDSE in 1D for the same atomic system 
used in the previous section (Section 4.1), but now the effective electric field is spatially
 inhomogeneous. Fig.~\ref{inHomoFieldVG_LG_CN} shows the comparison between 
 the calculated HHG spectra driven by an inhomogeneous field for both the LG and VG. 
 The inhomogeneous parameter value is set at $\epsilon=0.0175$~a.u., which correspond to an
 inhomogeneous region of about 60 a.u. (3 nm) (see~\cite{ciappi2012} for more details). Perfect agreement
 between the predictions of both the LG and VG are found. Therefore, these results suggest 
 that our derivations are appropriate for spatial nonhomogeneous fields as well. 
As a consequence, this invariance allows us to check 
which gauge can be more convenient to compute the 
HHG driven by spatial nonhomogeneous fields. We will address this point by considering the convergence of both the LG and VG. In other words, which of the two gauges presents less numerical error against the grid step, $\delta x$, and which one is faster in the computation of the HHG spectra. 

\begin{figure}
\begin{center}
\resizebox{1\textwidth}{!}{\includegraphics{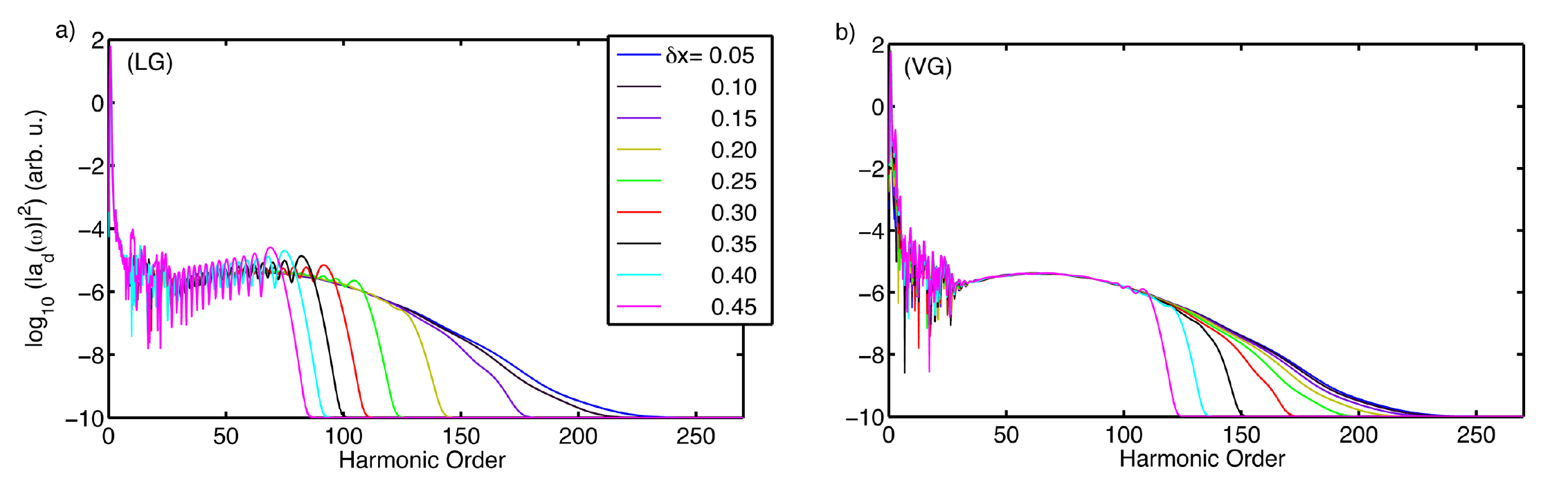}}
\caption{\label{InHomoField_SetData_VG_LG_CN} (color online) HHG spectra 
driven by plasmonic fields both in the LG and VG as a function of the grid
 step $\delta x$ are depicted in panel (a) and (b), respectively. We have used the 
 CN method to numerically integrate the TDSE.  The parameters for the laser 
 pulse are the same that those used in Fig.~\ref{inHomoFieldVG_LG_CN} 
 and the inhomogeneous parameter is $\epsilon=0.0175$ a.u.}
\end{center}
\end{figure}

\noindent Fig.~\ref{InHomoField_SetData_VG_LG_CN} shows the HHG
spectra as a function of the grid step for
 both the LG Fig.~4(a) and the VG Fig.~4(b) computed by using the CN method. The HHG spectra for the LG
 show a convergence for the smallest grid step, i.e., for $\delta x=0.05$ a.u. We should note, however, that
 the highest frequency of the HHG spectra change when the grid step is increases, which suggests that the
 computation of the HHG spectra driven by spatial inhomogeneous fields deserves special attention when``large" grid steps are used. A similar
 result is found when the VG is employed although it is possible to observe convergence for 
 larger values of $\delta x$. A suitable way to confirm the HHG cutoff and corroborate the convergence of the numerical schemes, is to rely on classical simulations. It is known that the limits on the HHG spectra can be obtained via classical simulations, e.g.~by computing the maximum electron kinetic energy upon recombination~\cite{maciej}. For spatial nonhomogeneous fields, it was demonstrated a perfect agreement between the classical predictions and the TDSE simulations (see e.g.~\cite{ciappi2012}) and, as a consequence, we could benchmark our VG and LG approaches by solving the classical equations of motion for an electron moving in an oscillating and spatial dependent electric field (for more details see~\cite{ciappicpc}).
 
\begin{figure}
\begin{center}
\resizebox{1.\textwidth}{!}{\includegraphics{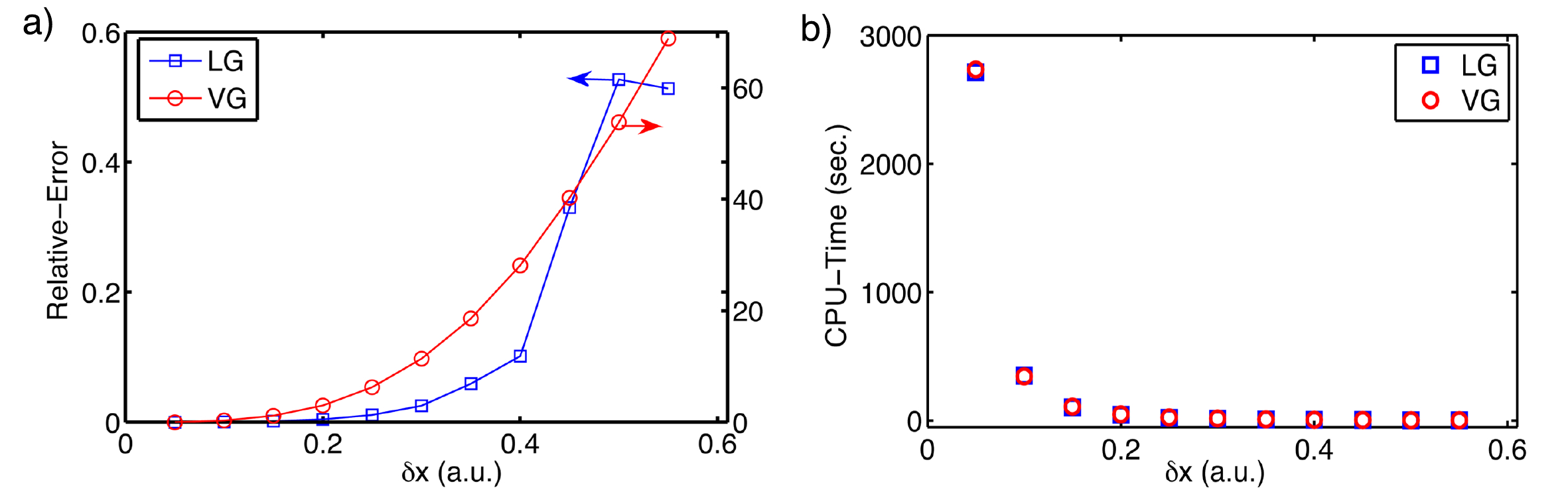}}
\caption{\label{NumConvergenceVG_LG_CN} (color online) (a) Convergence relative-error as a function of the
grid step by using the LG (blue line with squares) and the VG (red line with circles). 
(b) computing real-time as a function of the grid step for both the LG 
(blue squares) and VG (red circles). The simulation parameters are the same that those used in Fig.~\ref{InHomoField_SetData_VG_LG_CN}.}
\end{center}
\end{figure}

In addition, despite of the fact that for larger grid steps the LG shows a large 
deviation for the highest frequency compared 
to the VG results, we evaluate the relative error defined 
by $\frac{\Delta I_{\rm HHG,\delta x_j}}{I_{\rm HHG,\delta x_0}}$ 
for each gauge as a function of the grid step~$\delta x$. The results 
are depicted in Fig.~\ref{NumConvergenceVG_LG_CN}(a). This panel 
shows that a large difference appears whether LG or VG is used to compute 
the HHG spectra by the CN method. For values of $\delta x$ larger than $0.2$~a.u., 
the relative error between the LG and the VG has a difference of about two orders 
of magnitude, which suggests that the LG would be more appropriate than the 
VG to compute the HHG spectra.
Finally, in the panel (b), we show the computational time for each gauge. As can be observed the computing times for both the LG and the VG are similar. From this consideration we can conclude that the LG could be the most appropriated gauge in order to compute the HHG, given the fact it allows us to use larger grid steps.

\section{Conclusions}
We have reviewed the gauge invariance problem,  both analytical and numerically, for the calculation of the HHG phenomenon driven by spatial homogeneous and inhomogeneous (plasmonic enhanced) electric fields. To this purpose we have solved the TDSE in reduced dimensions by implementing the Spectral-Split Operator and the Crank-Nicolson algorithms. It was found that both the LG and VG are equivalent in the description of the harmonic emission processes for each of the two studied cases: the spatial homogeneous and the spatial inhomogeneous fields.  For the spatial inhomogeneous field case, and due to the dependence of the vector potential on the position, we found that the SO method was difficult to implement in the numerical solution of the TDSE. In contrast, the CN method has shown advantages because it is based on a finite element discretization. Our numerical results based on the CN method suggested that the calculation of the harmonic spectra depends strongly on the grid step chosen to perform the numerical integration. Both gauges are equivalent, but according to the numerical convergence of the HHG spectra, the LG apparently appears to be more accurate than the VG for the lowest harmonics. This is so because the lowest harmonics change by several orders of magnitude when the grid step increases. Furthermore, it was shown that particular attention in the choice of the spatial grid step should to be taken when spatial inhomogeneous fields are employed. This is so, because the limits of the harmonic radiation appear to be very sensitive to this parameter. Finally, it was found that the computational time was similar for both  the LG or VG, if they were used for the computation of the HHG spectrum in the moderate and high laser intensity regimes.

\section{Acknowledgments}
We are grateful to Alejandro De La Calle for the insightful discussion
and useful suggestions. A.C. and M.L. thanks the Spanish Ministry Project
FrOntiers of QUantum Sciences (FOQUS, FIS2013-46768-P) and ERC AdG OSYRIS  for
financial support. 

We also acknowledge the support from the
ERCs Seventh Framework Programme LASERLAB-EUROPE
III (grant agreement 284464) and the Ministerio de Econom\'ia y
Competitividad of Spain (FURIAM project FIS2013-47741-R).

\end{document}